\documentstyle[11pt]{article}
\newcommand{\lvec}[1]{|#1\!\!>}
\newcommand{\rvec}[1]{<\!\!#1|}
\newcommand{\ud}{{\mathrm{d}}}
\input epsf

\begin{document}
\begin{center}{\Large Susceptibility and Group Velocity in a Fully Quantized Model
For Electromagnetically Induced Transparency}\end{center}
\centerline{Xiong-Jun Liu\footnote{email:x.j.liu@eyou.com}  and
Mo-Lin Ge} \vspace{2mm}

\begin{center}\it Theoretical Physics
Division, Nankai Institute of Mathematics,Nankai
University\end{center}
\begin{center}\it Liuhui Center for Applied
Mathematics, Nankai University and Tianjin University, Tianjin
300071, P.R.China\end{center}
\begin{abstract}
We have developed a fully quantized model for EIT in which the
decay rates are taken into account. In this model, the general
form of the susceptibility and group velocity of the probe laser
we obtained are operators. Their expectation value and fluctuation
can be obtained on the Fock space. Furthermore the uncertainty of
the group velocity under very weak intensity of the controlling
laser and the uncertainty relation between the phase operator of
coupling laser and the group velocity are approximately given.
Considering the decay rates of various levels, we may analyze the
probe laser near resonance in detail and calculate the fluctuation
in both absorption and dispersion. We also discuss how the fully
quantized model reduces to a semiclassical model when the mean
photon
numbers of the coupling laser is getting large.
\end{abstract}
\section{Introduction} \indent

Controlling the phase coherence in ensembles of multilevel atoms
has led to the observation of many striking phenomena in the
propagation of near-resonant light. The notable examples include
ultraslow light pulse propagation, light storage, lasing without
inversion and EIT\cite{1,2,3,4,5,6}. Among these striking
phenomena, ultraslow light speed and superluminality look more
attracting. The drastic reductions in the group velocity of pulses
were discussed in Ref.\cite{5} and recent experiments have taken
the reduction of the speed of light to extreme limits (shown in
Refs.\cite{6,7} and even to zero\cite{8,9}. On the other way, the
superluminal velocity was also observed in the abnormal-dispersion
media\cite{10}. In the most previous work the phenomenon of EIT
and the accompanying enhancement of the index of refraction and
susceptibility are treated using semiclassical theory in which
both the coupling and probe lasers were treated as classical. In
such a treatment, the occurrence of EIT requires the coupling
laser to be much stronger than the probe laser. In Ref.\cite{8},
Fleischhauer and Lukin treated the probe laser as quantized, and
showed that the quantum description of laser is more fundamental
than the classical one, having advantages in uncovering new
effects of EIT. In Ref.\cite{11} Kuang et al. treated both the
coupling and probe lasers as quantized and they pointed out that
in general the group velocity depends on the intensity of the
coupling as well as probe laser. But they ignored decay rates of
various levels, so their treatment is essentially a
time-independent approach. Obviously, if the decay rates are
incorporated, it needs to solve the evolution equation of the
density matrix, which is difficult in a fully quantized treatment.

Based on the results of the previous works, the strength of the
coupling laser needs to be modified from finite to zero to
decelerate and stop the input pulse\cite{8,9}. But when coupling
laser is very weak, we can no longer treat the coupling laser as
classical, in other words, we should develop a fully quantized
model for EIT.

In this paper, both the coupling and probe lasers are treated as
quantized. First we give a straightforward discussion of the case
with no decay rates In section 2.1, and then take the decay rates
into account in section 2.2. We shall obtain the general form of
density matrix and susceptibility and analyze several cases in
section 2.2: In part(a) we discuss how the fully quantized model
reduces to the model given by Refs\cite{8,9} when the mean photon
number of the coupling laser is large; Part(b) we obtain that the
general form of susceptibility and group velocity is operators,
the expectation value of which can be obtained by act the
operators on the Fock space and there is fluctuation. We calculate
the uncertainty of group velocity numerically and give an
approximate uncertainty relation between the phase operator of
coupling laser and the group velocity; Part(c) we discuss the more
general case where both the probe and coupling lasers are weak and
have similar intensities.

\section{The theoretical model}
\subsection{The case with no decay rates}

Let us start from the well-known three-level $\Lambda$-type
configuration atom (Fig.1) whose energy levels assumed to be
$E_a>E_c>E_b$. They interact with two quantized fields, probe and
coupling ones. The two low levels $\lvec{b}$ and $\lvec{c}$ are
coupled to the upper one $\lvec{a}$ separately and initially the
atom is in the ground state $\lvec{b}$. The frequency of the
coupling laser $\omega_2=\omega_{ac}$ and the probe laser
$\omega_1=\omega_{ab}-\Delta_1$, where $\Delta_1$ is the detuning
of the probe laser. Here both the probe and coupling lasers are
quantized. In the interaction picture the Hamiltonian of the
system is\cite{12}:
\begin{eqnarray}
\label{eqn:1}  H^I&=&H_0^I+H_1^I
\\ \nonumber\\
\label{eqn:2}
H_0^I&=&E_a\lvec{a}\rvec{a}+E_b\lvec{b}\rvec{b}+E_c\lvec{c}\rvec{c}
\nonumber\\
&&-\hbar[\Delta_1(\lvec{a}\rvec{a}+\lvec{c}\rvec{c})]
+\omega_1\hat{a}_1^{\dagger}\hat{a}_1+
\omega_2\hat{a}_2^{\dagger}\hat{a}_2 \\
\nonumber \\
 \label{eqn:3}
H_1^I&=&\hbar\Delta_1(\lvec{a}\rvec{a}+\lvec{c}\rvec{c})
-\hbar(g_1\hat{a}_1 \lvec{a}\rvec{b}+g_2\hat{a}_2
\lvec{a}\rvec{c}+\rm{H.C})
\end{eqnarray}
where $\hat{a}_i$ and $\hat{a}_i^{\dagger}$ ($i=1,2$) are the
annihilation and creation operators of the probe (for $i=1$) and
coupling (for $i=2$) laser modes respectively, and $g_i$ the
coupling constants. Assuming the detuning $\Delta_1$ small, we
shall find the solution for the dark-state $\lvec{\psi_0}$ of the
system by perturbative approximation. The perturbation with the
first order was given in Ref.\cite{11}:
\begin{eqnarray}
\label{eqn:4}
\lvec{\psi_0}&=&\lvec{\psi_0^{(0)}}+\lvec{\psi_0^{(1)}}\nonumber\\
&=&\frac{\Omega_2}{\Omega}\lvec{b,n_1,n_2}- \frac{2\Omega_1
\Omega_2}{\Omega^3}\Delta_1\lvec{a,n_1-1,n_2}-
\frac{\Omega_1}{\Omega}\lvec{c,n_1-1,n_2+1}
\end{eqnarray}
With the energy eigenvalue
$\hbar\frac{\Omega_1^2}{\Omega^2}\Delta_1$, i.e.,
$H_1^I\lvec{\psi_0}=\hbar\frac{\Omega_1^2}{\Omega^2}\Delta_1
\lvec{\psi_0}$. The second-ordered perturbation can be calculated:
\begin{eqnarray}
\lvec{\psi_0}&=&\lvec{\psi_0^0}+\lvec{\psi_0^1}+
\lvec{\psi_0^2}\nonumber\\
&=&\left(\frac{\Omega_2}{\Omega}-\frac{4\Omega_1^2\Omega_2^3}{\Omega^7}
\Delta_1^2\right)\lvec{b,n_1,n_2} -\frac{2\Omega_1
\Omega_2}{\Omega^3}\Delta_1\lvec{a,n_1-1,n_2}\nonumber\\
&&-\left(\frac{\Omega_1}{\Omega}+\frac{4\Omega_1\Omega_2^4}{\Omega^7}
\Delta_1^2\right) \lvec{c,n_1-1,n_2+1}\label{eqn:5}
\end{eqnarray}
where $\lvec{n_1,n_2}$ is the usual two-mode Fock basis.
$\Omega_1=2g_1\sqrt{n_1}$ , $\Omega_2=2g_2\sqrt{n_2+1}$ and
$\Omega=\sqrt{\Omega_1^2+\Omega_2^2}$. Suppose the coupling and
probe lasers are in a two-mode coherent state
$\lvec{\alpha,\beta}$ with$\alpha,\beta$ the real for
simplicity[12] and the atom is initially in the ground state
$\lvec{b}\!\!\otimes\lvec{\alpha,\beta}$. If we consider the ideal
case in which the decay rates of various levels are ignored and
initially $\Omega_1=0$ with $\Omega_2$ finite, then proceed to
turn $\Omega_2$ down while slowly turning $\Omega_1$ on. During
the course, the state $\lvec{\psi_0(t)}$ of the system will evolve
adiabatically, so from (\ref{eqn:4}) or (\ref{eqn:5}) we can get
the density matrix of the system. Further we can calculate the
susceptibility of the system. If the mean photon number of the
coupling laser $\bar{n}_{\alpha}=\alpha^2$ and the probe laser
$\bar{n}_{\beta}=\beta^2$ are large, i.e., in semiclassical limit,
from (\ref{eqn:5}) we have
\begin{eqnarray}\label{eqn:6}
\rho_{ab}(\omega_1)=
-\frac{2\bar{\Omega}_1\bar{\Omega}_2^2}{(\bar{\Omega}_1^2+\bar{\Omega}_2^2)^2}\Delta_1
+\frac{8\bar{\Omega}_1^3\bar{\Omega}_2^4}{(\bar{\Omega}_1^2+\bar{\Omega}_2^2)^5}\Delta_1^3
\end{eqnarray}
where $\bar{\Omega}_i=\bar{\Omega}_i(\bar{n}_{\alpha},
\bar{n}_{\beta})\,(i=1,2)$ are the Rabi frequencies of the
coupling and probe lasers, with
$\wp_{ab}N\rho_{ab}(\omega_1)=\epsilon_0\chi(\omega_1)
\hat{E}_1(\omega_1)$, The susceptibility is given by:
\begin{eqnarray}\label{eqn:7}
\chi(\omega_1)=-\frac{4N|\wp_{ab}|^2\bar{\Omega}_2^2}
{\hbar\epsilon_0(\bar{\Omega}_1^2+\bar{\Omega}_2^2)^2}\Delta_1
+\frac{16|\wp_{ab}|^2N\bar{\Omega}_1^2\bar{\Omega}_2^4}
{\hbar\epsilon_0(\bar{\Omega}_1^2+\bar{\Omega}_2^2)^5}\Delta_1^3
\end{eqnarray}
and
\begin{eqnarray}\label{eqn:8}
\frac{\ud \chi}{\ud
\omega_1}=\frac{4N|\wp_{ab}|^2\bar{\Omega}_2^2}
{\hbar\epsilon_0(\bar{\Omega}_1^2+\bar{\Omega}_2^2)^2}
-\frac{48|\wp_{ab}|^2N\bar{\Omega}_1^2\bar{\Omega}_2^4}
{\hbar\epsilon_0(\bar{\Omega}_1^2+\bar{\Omega}_2^2)^5}\Delta_1^2
\end{eqnarray}

The above result is valid for small $\Delta_1$.The last term of
r.h.s of (\ref{eqn:8}) which was not included in the Ref.\cite{11}
is negative and it indicates that $\frac{\ud \chi(\omega_1)}{\ud
\omega_1}<0$ and then group velocity of the probe laser may be
greater than the vacuum speed c if we extrapolate (\ref{eqn:8})to
large $\Delta_1$ for the abnormal dispersion we meet here.

The case we discussed above is very ideal. However, in general the
decay rates of various levels cannot be ignored. In this case, to
obtain the susceptibility of the media, we should solve the
evolution equation of the density matrix.

\subsection{The case with decay rates}
In the Schr\"{o}dinger picture, the dynamics of the system is
described by the interaction Hamiltonian:
\begin{eqnarray}\label{eqn:9}
H_I&=&-N\int \frac{\ud z}{L}({\hbar}g_1\hat{a}_1
e^{i\frac{\omega_1}{c}(z-ct)}|a><b| \nonumber \\
&&+{\hbar}g_2\hat{a}_2e^{i\frac{\omega_2}{c}(z-ct)}|a><c|
+\rm{H.c})
\end{eqnarray}
where $\omega_1=\omega_{ab}-\Delta_1,\,\omega_2=\omega_{ac},
\,g=\wp\sqrt{\frac{\omega}{2\hbar\epsilon_0V}}$and $V$ the
quantization volume, $N$ the number of atoms in this volume and
$L$ its length in $z$ direction. The density matrix of the atom
system is defined by:
\begin{eqnarray}\label{eqn:10}
\rho(z,t,t_0)=\sum_{\alpha,\:\beta} \rho_{\alpha,\beta}(z,t,t_0)
\lvec{\alpha}\rvec{\beta}
\end{eqnarray}
where $\alpha,\:\beta=a,\:b,\:c$, and $\rho_{\alpha\beta}$ are the
density matrix elements. Make the substitutions:
$\rho_{ab}=\tilde{\rho}_{ab}e^{-i(\omega_{ab}-\Delta_1)t},
\:\rho_{cb}=\tilde{\rho}_{cb}e^{-i(\omega_{cb}-\Delta_1)t}$, and
others,
$\rho_{\mu\nu}=\tilde{\rho}_{\mu\nu}e^{-i\omega_{\mu\nu}t}$. If
the initial state of the atom-field system is assumed to be
$\lvec{b}\!\otimes\!\lvec{\alpha,\beta}$, i.e.,
$\bar{\Omega}_1(t=0)=0,\,\bar{\Omega}_2(t=0)>0$. In the
near-resonance case, very little atoms are populated in the state
$\lvec{a}$ and the matrix elements $\tilde{\rho}_{bb}$
and$\tilde{\rho}_{cc}$ varies slowly with $t$, therefore these two
density matrix elements can be, respectively, replaced by their
initial value $\tilde{\rho}_{bb}^{(0)}$ and
$\tilde{\rho}_{cc}^{(0)}$ which are given by (\ref{eqn:4}) or
(\ref{eqn:5}), then the evolution equations of the three density
matrix elements $\tilde{\rho}_{ab},\tilde{\rho}_{cb}$ and
$\tilde{\rho}_{ca}$  can be written in the matrix form:
\begin{eqnarray}
\dot{R}=-MR+A\label{eqn:11}
\end{eqnarray}
where
\begin{eqnarray}
\label{eqn:12} &&R= \left[ \matrix{\tilde{\rho}_{ab}\cr
\tilde{\rho}_{cb}\cr \tilde{\rho}_{ca}}\right],  \quad M=\left[
\matrix{ \gamma_1+i\Delta_1 &-ig_2\hat{a}_2e^{ik_2z} &0\cr
-ig_2\hat{a}_2^{\dagger}e^{-ik_2z} &\gamma_3+i\Delta_1
&ig_1\hat{a}_1e^{ik_1z}\cr 0 &ig_1\hat{a}_1^{\dagger}e^{-ik_1z}
&\gamma_2} \right] ,\nonumber \\
 &&\;\;\;\;\;\; A=\left[\matrix{
i\frac{\wp_{ab}}{2\hbar}\tilde\rho_{bb}^{(0)}\hat{E}_1(z)\cr 0\cr
i\frac{\wp_{ac}}{2\hbar}\tilde\rho_{cc}^{(0)}\hat{E}_2(z)}\right]
\end{eqnarray}
and
$\hat{E}_m(z)=\sqrt{\frac{\hbar\omega_i}{2\epsilon_0V}}e^{ik_mz}\,
(m=1,2)$, $\gamma_1,\,\gamma_2$ and $\gamma_3$ are the
off-diagonal decay rates for
$\tilde{\rho}_{ab},\,\tilde{\rho}_{ca}$ and $\tilde{\rho}_{cb}$
respectively. Conventionally, both the coupling and probe lasers
were treated as classical, it should require the coupling laser is
much stronger than the probe laser, hence only $\tilde{\rho}_{ab}$
and $\tilde{\rho}_{cb}$ are needed\cite{4}. However in our case,
the coupling and probe laser are both quantized, so they may be
equally strong, therefore, besides $\tilde{\rho}_{ab}$ and
$\tilde{\rho}_{cb}$, we should consider $\tilde{\rho}_{ca}$ as
well.

When the matrix $M$ is non-singular, the formal solution of
equation (\ref{eqn:11}) is given by:
\begin{eqnarray}\label{eqn:13}
R=e^{-Mt}R_0+(1-e^{-Mt})M^{-1}A
\end{eqnarray}
where $R_0$ is the initial value of $R(t)$ given by (\ref{eqn:4})
or (\ref{eqn:5}). Let us first make analysis of solutions of
(\ref{eqn:13}) varying with $t$:

i) when $t$ is very small, i.e., $e^{-Mt}\sim 1$, so $R(t\to
0)=R_0$. However, because of the decay rates, only initially $R$
can be given by (\ref{eqn:4}) or (\ref{eqn:5}) and its form will
be changed when time becomes large.

ii) when $t$ is large, i.e., $e^{-Mt}\sim 0$, then
\begin{eqnarray}\label{eqn:14}
R=M^{-1}A
\end{eqnarray}
From (\ref{eqn:14}) we see that $R(t)$ will reach a steady value
when the time is large enough, i.e., $R$ become independent of
time. Under such condition we should carefully consider various
cases as follows:

(a) If the mean photon number of coupling laser
$\bar{n}_{\alpha}=\alpha^2$ is large while that of the probe laser
$\bar{n}_{\beta}=\beta^2$ is small, and
$\bar{n}_{\alpha}\gg\bar{n}_{\beta}$, therefore the influence of
the probe laser in equation (\ref{eqn:11}) can be ignored and most
atoms populated in the ground state, i.e.,
$\tilde{\rho}_{bb}^{(0)}=1, \tilde{\rho}_{cc}^{(0)}=0$, so the
matrix $M$ and $A$ are given by
\begin{equation}
 \quad M=\left[ \matrix{\gamma_1+i\Delta_1
&-ig_2\hat{a}_2e^{ik_2z} &0\cr -ig_2\hat{a}_2^{\dagger}e^{-ik_2z}
&\gamma_3+i\Delta_1 &0\cr 0 &0 &\gamma_2}\right], \quad A=\left[
\matrix{i\frac{\wp_{ab}}{2\hbar}\hat{E}_1(z)\cr 0\cr
0}\right]
\label{eqn:15}
\end{equation}
We then obtain
\begin{eqnarray}\label{eqn:16}
\tilde{\rho}_{ab}=\frac{i\frac{\wp_{ab}}{2\hbar}(\gamma_3+i\Delta_1)\hat{E}_1(z)}
{(\gamma_1+i\Delta_1)(\gamma_3+i\Delta_1)+g_2^2\hat{a}_2\hat{a}_2^{\dagger}}
\end{eqnarray}
Together with
$\wp_{ab}N\tilde{\rho}_{ab}=\epsilon_0\chi(\omega_1)\hat{E}_1(z)$,
the susceptibility is given by
\begin{eqnarray}\label{eqn:17}
\hat{\chi}(\omega_1)=\frac{ig_1^2N(\gamma_3+i\Delta_1)}
{\omega_1\left[(\gamma_1+i\Delta_1)(\gamma_3+i\Delta_1)+
g_2^2\hat{a}_2\hat{a}_2^{\dagger}\right]}
\end{eqnarray}
which is a formal solution and the susceptibility is an operator .
Because $\bar{n}_{\alpha}=\alpha^2$ is large and the relative
fluctuation of the photon number $\frac{\Delta
n}{\bar{n}_{\alpha}}=\frac{1}{\alpha}$ is small, we may, as a good
approximation, replace $g_2^2\hat{a}_2\hat{a}_2^{\dagger}$ by
$g_2^2(\bar{n}_{\alpha}+1)$ when calculating the mean value of
$\chi(\omega_1)$ in the two-mode coherent state
$\lvec{\alpha,\beta}$. Noting that
$\bar{\Omega}_2=g_2\sqrt{\bar{n}_{\alpha}+1}$, then
\begin{eqnarray}\label{eqn:18}
\bar{\chi}(\omega_1)=\frac{ig_1^2N(\gamma_3+i\Delta_1)}
{\omega_1\left[(\gamma_1+i\Delta_1)(\gamma_3+i\Delta_1)
+\bar{\Omega}_2^2\right]}
\end{eqnarray}

The above result is familiar\cite{4}, which indicates that when
the mean photon number of the coupling laser is large, our model
reduces to the model where the probe laser is quantized while the
coupling laser is classical\cite{8,9}.

(b) If both the probe laser and coupling laser are very weak, but
still $\bar{n}_{\alpha}\gg\bar{n}_{\beta}$, the formal solution of
the susceptibility has the same form as (\ref{eqn:17}), however,
meanwhile we should not ignore the relative fluctuation of photon
numbers of the coupling laser. The relative fluctuation of the
susceptibility may also be large, and
$g_2^2\hat{a}_2\hat{a}_2^{\dagger}$ can no longer be replaced by
$g_2^2(\bar{n}_{\alpha}+1)$ in the calculation of the mean value
of $\chi(\omega_1)$. In fact, we have in this case
\begin{eqnarray}\label{eqn:19}
\bar{\chi}(\omega_1)&=&\rvec{\alpha,\beta}
\hat{\chi}(\omega_1)\lvec{\alpha,\beta}
           \nonumber\\
&=&\sum_{n_2=0}^{\infty} \frac{ig_1^2N(\gamma_3+i\Delta_1)}
{\omega_1\left[(\gamma_1+i\Delta_1)(\gamma_3+i\Delta_1)+
g_2^2(n_2+1)\right]} \cdot\frac{\alpha^{2n_2}}{n_2!} e^{-\alpha^2}
\end{eqnarray}

Noting that $\bar{\chi}(\omega_1)=\bar{\chi}_1(\omega_1)
+i\bar{\chi}_2(\omega_1)$, $\bar{\chi}_1(\omega_1)$ and
$\bar{\chi}_2(\omega_2)$ are, respectively, the real and imaginary
parts of the complex susceptibility and related to the dispersion
and absorption:
\begin{eqnarray}\label{eqn:20}
\bar{\chi}_1(\omega_1) =g_1^2Ne^{-\alpha^2} \sum_{n_2=0}^{\infty}
\frac{\left(\gamma_3+\Delta_1^2-g_2^2(n_2+1)\right)\Delta_1}
{\omega_1\left[\left(\gamma_1\gamma_3-\Delta_1^2+g_2^2(n_2+1)\right)^2
+(\gamma_1+\gamma_3)^2\Delta_1^2\right]}
\frac{\alpha^{2n_2}}{n_2!}
\end{eqnarray}
and
\begin{eqnarray}\label{eqn:21}
\bar{\chi}_2(\omega_1)=g_1^2Ne^{-\alpha^2} \sum_{n_2=0}^{\infty}
\frac{\left(\gamma_1\gamma_3^2+g_2^2(n_2+1)\right)\Delta_1}
{\omega_1\left[\left(\gamma_1\gamma_3-\Delta_1^2+g_2^2(n_2+1)\right)^2
+(\gamma_1+\gamma_3)^2\Delta_1^2\right]}
\frac{\alpha^{2n_2}}{n_2!}
\end{eqnarray}
Noting that
$P_1=\Delta\chi_1(\omega_1)/{|\bar{\chi}_1(\omega_1)|}$ and
$P_2=\Delta\chi_2(\omega_1)/{|\bar{\chi}_2(\omega_1)|}$ are the
relative fluctuation of $\chi_1(\omega_1)$ and $\chi_2(\omega_1)$
respectively. in Fig.2 and Fig.3, the relative fluctuation of
$\chi_1(\omega_1)$ and $\chi_2(\omega_1)$ are plotted versus the
detuning $\Delta_1$ in units of the atomic decay $\gamma_1$
respectively, for $\alpha^2=500$, $\gamma_1\gg\gamma_3$ and
$\bar{\Omega}_2=g_2\sqrt{\bar{n}_{\alpha}+1}=\gamma_1/2$. It is
seen that, around the zero detuning, for example,
$\Delta_1\approx-0.1\gamma_1$. The relative fluctuation of
$\chi_1(\omega_1)$ is small ($\approx 4\%$), while that of
$\chi_2(\omega_1)$ is large ($\approx 9\%$). On the other hand,
around the detuning $\Delta_1=-0.7$ the relative fluctuation of
$\chi_1(\omega_1)$ is large ($\approx 200\%$), while that of
$\chi_2(\omega_1)$ is small ($\approx 0.4\%$). Furthermore, the
derivative of $\hat{\chi}_1(\omega_1)$ is related to the group
velocity for the probe laser pulse through
\begin{eqnarray}\label{eqn:22}
\hat{V}_g=c/{\left[1+(\omega_1/2)(\frac{\ud \hat{\chi}_1}{\ud
\omega_1})\right]}
\end{eqnarray}
from which we can further numerically calculate the accompany
fluctuation of the velocity. For example, on the zero detuning
$\Delta_1=0$, we obtain $\bar{V}_g=10.02$ m/s, uncertainty $\Delta
V_g=0.45$ m/s, and relative fluctuation $\Delta
V_g/\bar{V}_g\approx 4.5\%$, while on the detuning
$\Delta_1=0.16\gamma_1$, we obtain$\bar{V}_g=39.45$ m/s,
uncertainty $\Delta V_g=27.90$ m/s, and relative fluctuation
$\Delta V_g/\bar{V}_g\approx 70.7\%$.

The results above shows that the group velocity of the probe laser
is not a certainty in the fully quantized model. Its uncertainty
is a function of detuning $\Delta_1$. In what follows we shall
give an approximate uncertainty relation between the phase
operator of coupling laser and the group velocity, for the
two-mode coherent state $\lvec{\alpha,\beta}$, as an
approximation, we have
\begin{eqnarray}\label{eqn:23}
\hat{V}_g \approx \bar{V}_g(\bar{n}_{\alpha},\Delta_1)+
\frac{\partial{f(n_2,\Delta_1)}}{\partial{n_2}}|_{n_2=n_{\alpha}}(\hat{n}_2-\bar{n}_\alpha)
\end{eqnarray}
where $f(n_2,\Delta_1)=c/{[1+(1/2)\frac{\ud}{\ud\omega_1}
(\frac{g_1^2N(\Delta_1^3-g_2^2(n_2+1)\Delta_1)}
{(g_2^2(n_2+1)-\Delta_1^2)^2 +\gamma_1^2\Delta_1^2})]}$, and
$\hat{n}_2=\hat{a}_2^{\dagger}\hat{a}_2$ is the particle number
operators of the coupling laser modes and it satisfies the
commutation relation\cite{14}:
\begin{eqnarray}\label{eqn:24}
[\hat{n}_2,\cos{\hat{\phi}}]=-i\sin{\hat{\phi}}
\end{eqnarray}
where $\cos{\hat{\phi}}$ is the phase operator of the coupling
laser, then
\begin{eqnarray}\label{eqn:25}
[\hat{V}_g,\cos{\hat{\phi}}]=-i
F(n_\alpha,\Delta_1)\sin{\hat{\phi}}
\end{eqnarray}
where $F(n_\alpha,\Delta_1)=\frac{\partial{f(n_2,\Delta_1)}}
{\partial{n_2}}|_{n_2=n_\alpha}$, from the uncertainty principle,
we have
\begin{eqnarray}\label{eqn:26}
\langle\Delta\hat{V}_g\rangle\langle\Delta\cos{\hat{\phi}}\rangle\geq
\frac{1}{2}|F(n_\alpha,\Delta_1)\langle\sin{\hat{\phi}}\rangle|
\end{eqnarray}
which turns out that the uncertainty of the group velocity is the
function of $n_\alpha$ and $\Delta_1$.

As we have known that to decelerate and stop the input pulse, the
strength of the coupling laser need to be modified from finite to
zero \cite{8,9,13}, but when $\Omega_2$ is very small, we should
treat the coupling laser as quantized. It is noticeable that if
initially the coupling laser is much strong than the probe one,
the coupling laser will be much stronger than the probe laser at
all times(see Ref.\cite{8}), which satisfies the condition of part
(b) we discussed above.

(c) When both the probe and coupling lasers are very weak and have
similar intensities, i.e., $\bar{n}_{\alpha}\sim\bar{n}_{\beta}$,
from (\ref{eqn:4}) we have
$\tilde{\rho}_{bb}^{(0)}=\frac{\Omega_2^2}{\Omega^2}$ and
$\tilde{\rho}_{cc}^{(0)}=\frac{\Omega_1^2}{\Omega^2}$, hence from
the equation (\ref{eqn:11}) the form of the density matrix element
$\tilde{\rho}_{ab}$ can be followed:
\begin{eqnarray}\label{eqn:27}
\tilde{\rho}_{ab}=
\frac{i\frac{\wp_{ab}}{2\hbar}\hat{E}_1(z)\rho_{bb}^{(0)}
\left[(\gamma_3+i\Delta_1)\gamma_2+g_1^2\hat{a}_1\hat{a}_1^{\dagger}\right]}
{(\gamma_1+i\Delta_1)(\gamma_3+i\Delta_1)\gamma_2+
\gamma_2g_2^2\hat{a}_2\hat{a}_2^{\dagger}
+(\gamma_1+i\Delta_1)g_1^2\hat{a}_1\hat{a}_1^{\dagger}}
\end{eqnarray}
and
\begin{eqnarray}\label{eqn:28}
\hat{\chi}(\omega_1)= \frac{ig_1^2N\rho_{bb}^{(0)}
\left[(\gamma_3+i\Delta_1)\gamma_2+g_1^2\hat{a}_1\hat{a}_1^{\dagger}\right]}
{\omega_1\left[(\gamma_1+i\Delta_1)(\gamma_3+i\Delta_1)\gamma_2+
\gamma_2g_2^2\hat{a}_2\hat{a}_2^{\dagger}
+(\gamma_1+i\Delta_1)g_1^2\hat{a}_1\hat{a}_1^{\dagger}\right]}
\end{eqnarray}
From the above result we find that the susceptibility depends on
the coupling laser as well as the probe laser when both of them
are equally strong, which is similar to the result of KCW in
Ref\cite{11}, but here the susceptibility is an operator and its
value can be calculated on the Fock space just as the analysis in
part(b).

\section{Conclusion}

We have developed a fully quantized model for EIT in which the
decay rates are taken into account. In this model, to solve the
evolution equation for density matrix, we separate the atom-system
from the photon-system described by Fock-states and only calculate
the density matrix of atom-system. By this means, the general form
of susceptibility and group velocity of the probe laser we
obtained are operators concerned with particle number operators of
the probe and coupling laser modes. Their expectation value and
fluctuation can be calculated on the Fock space. We have
calculated the uncertainty of the group velocity numerically and
give an approximate uncertainty relation between the phase
operator of the coupling laser and the group velocity. When both
the probe and coupling lasers are weak and have similar
intensities, we find the susceptibility depends on the coupling
laser as well as the probe laser. Considering the decay rates of
various levels, we can make analysis of absorption of probe laser
near resonance and calculate the fluctuation in both absorption
and dispersion. We also discuss how the fully quantized model
reduces to a semiclassical model when the mean photon numbers of
the coupling laser is large.

\vspace{5mm} This work is in part supported by NSF of China.

\newpage
\begin{figure}[ht]
\epsfsize=0.8\columnwidth\epsfbox{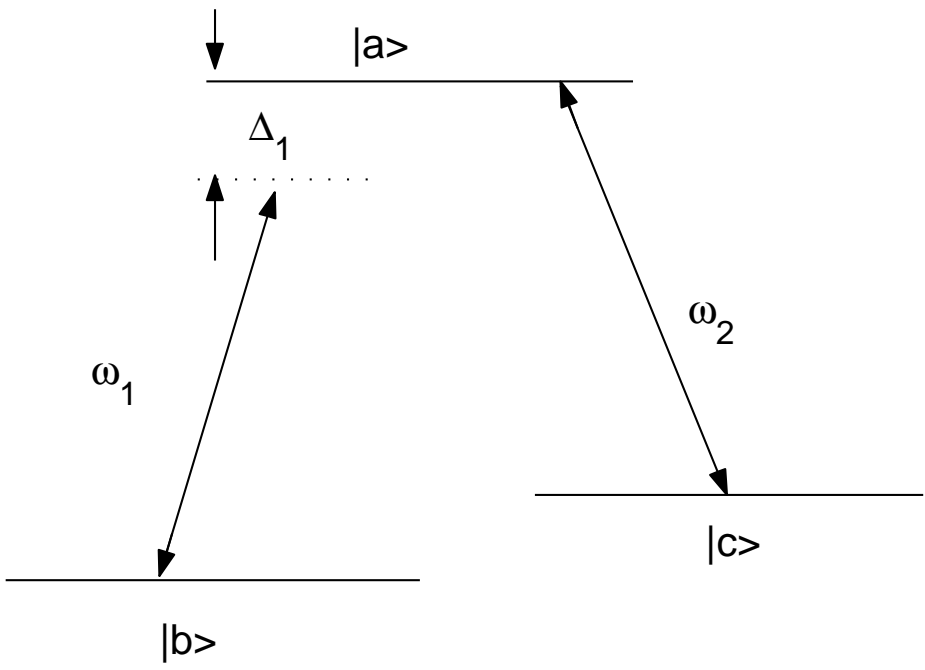} \caption{Energy levels
of a $\Lambda$-type atom}\label{}
\end{figure}
\begin{figure}[ht]
\epsfsize=0.8\columnwidth \epsfbox{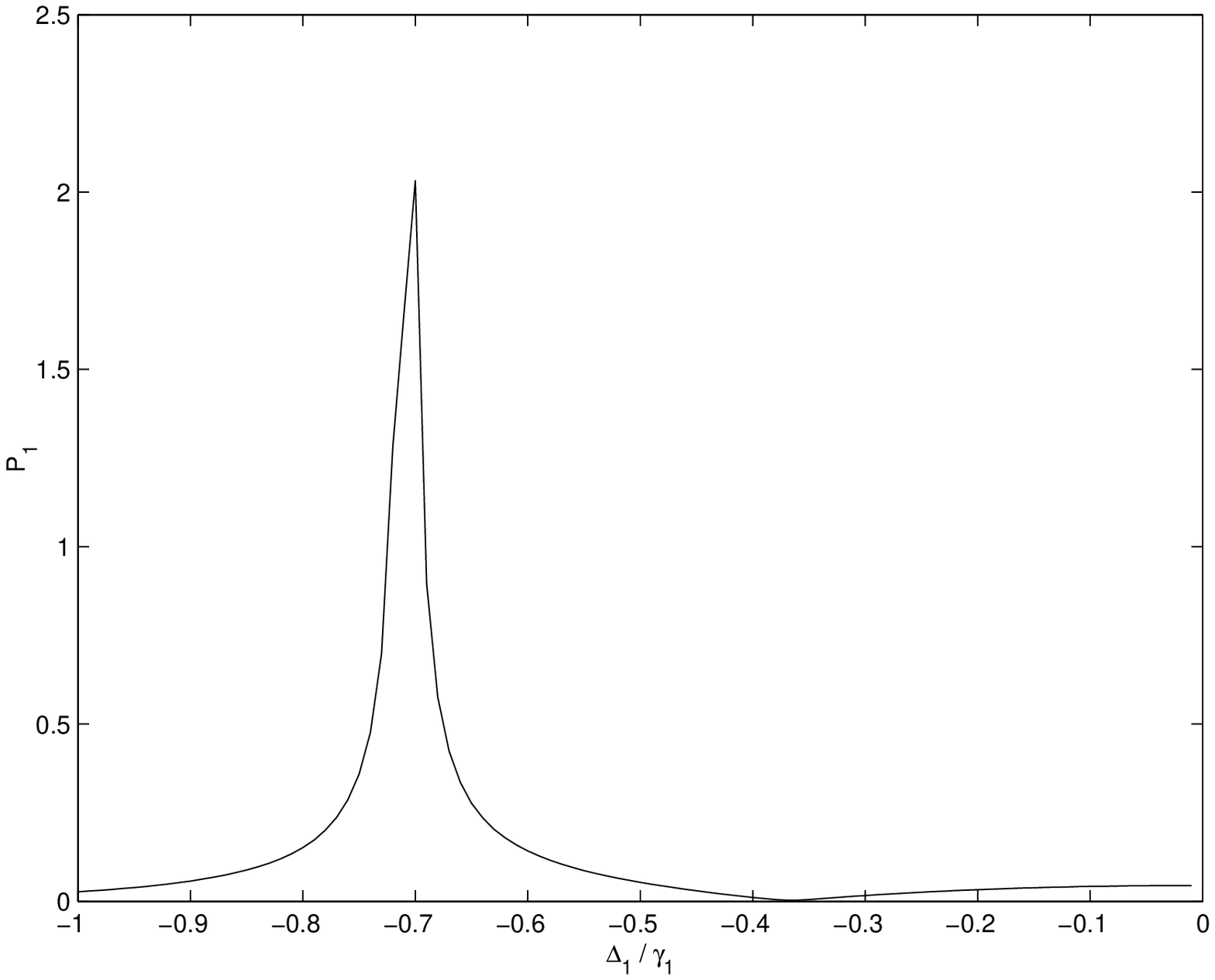} \caption{the relative
fluctuation of $\chi_1(\omega_1)$, for $\alpha^2=500$,
$\gamma_1\gg\gamma_3$ and
$\bar{\Omega}_2=g_2\sqrt{\bar{n}_{\alpha}+1}=\gamma_1/2$.}\label{}
\end{figure}
\begin{figure}[ht]
\epsfsize=0.8\columnwidth \epsfbox{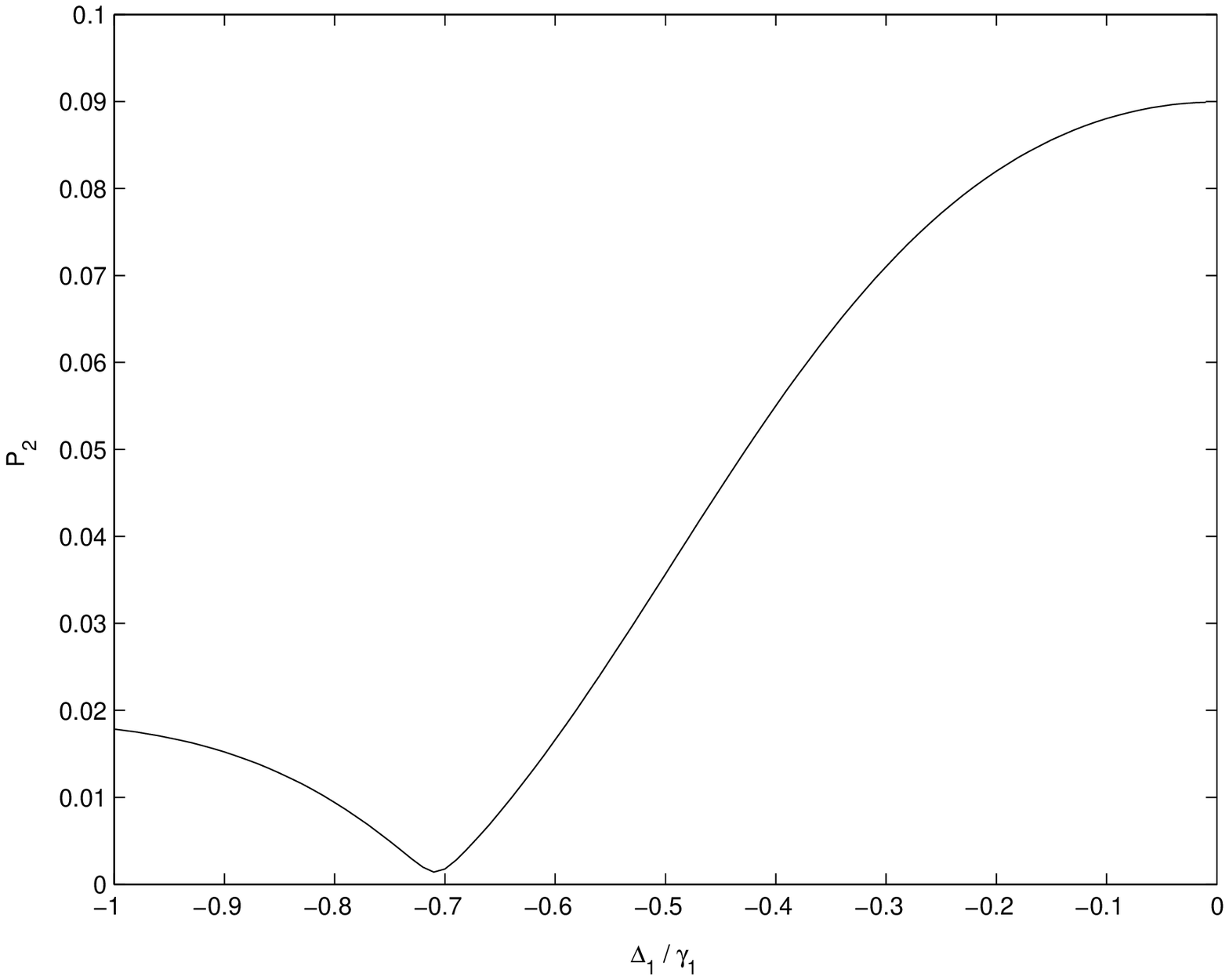} \caption{the relative
fluctuation of $\chi_2(\omega_1)$, for $\alpha^2=500$,
$\gamma_1\gg\gamma_3$ and
$\bar{\Omega}_2=g_2\sqrt{\bar{n}_{\alpha}+1}=\gamma_1/2$.}\label{}
\end{figure}


\begin{thebibliography}{99}
\bibitem{1} L.V.Hau et al., Nature (London) 397,594(1999)
\bibitem{2} M.M.Kash et al., Phys.Rev.Lett.82, 529(1999)
\bibitem{3} C.Liu, Z.Dutton, C.H.Behroozi, and L.V.Hau et al., Nature (London) 409,490(2001)
\bibitem{4} M.O.Scully and M.S.Zubairy, Quantum Optics (Cambridge University Press, Cambridge, 1999)
\bibitem{5} S.E.Harris, J.E.Field, and A.Kasapi, Phys.Rev.A, 46,R29(1992)
\bibitem{6} M.D.Lukin et al. Phys.Rev.Lett. 79,2959(1997)
\bibitem{7} D.Budker et al. Phys.Rev.Lett. 83,1767(1999)
\bibitem{8} M.Fleischhauer and M.D.Lukin, Phys.Rev.Lett. 84, 5094 (2000)
\bibitem{9} M.Fleischhauer and M.D.Lukin, Phys.Rev.A 65,022314(2002)
\bibitem{10} A.Dogariu, A.Kuzmich, and L.J.Wang, Phys.Rev.A 63,053806(2001)
\bibitem{11} Le-Man Kuang, Guang-Hong Chen, and Yong-Shi Wu, arXiv: quantum-ph/0103152(2001)
\bibitem{12} A.B.Klimov, L.L.Sanchez-Soto, J.Delgado, and E.C.Yustas, Phys.Rev.A 67,013803(2003)
\bibitem{13} D.F.Phillips, A.Fleischhauer et al., Phys.Rev.Lett.86,783(2001)
\bibitem{14} Masanao Ozawa, Annals Phys. 257,65(1997); E-print: quant-ph/9705034
\end{thebibliography}
\end{document}